\pdfoutput=1
\documentclass[prepreint,preprintnumbers,amsmath,amssymb,floatfix]{revtex4}
\usepackage[pdftex]{graphicx}
\usepackage{color}
\usepackage{dcolumn}
\usepackage{bm}
\usepackage{subfigure}
\usepackage{times}
\usepackage{amsmath}
\usepackage{amssymb}
\usepackage{appendix}
\newcommand\nn{\nonumber}
\newcommand\bea{\begin{eqnarray}}
\newcommand\eea{\end{eqnarray}}
\newcommand\f{\frac}
\newcommand\p{\partial}
\newcommand\la{\langle}
\newcommand\ra{\rangle}

\newcommand\ie{{\emph{i.e.}}}

\begin{document}

\title{Numerical test of hydrodynamic fluctuation theory in the  Fermi-Pasta-Ulam chain}
\author{Suman G. Das}
\email{suman@rri.res.in}
\affiliation{Raman Research Institute, CV Raman Avenue, Sadashivanagar, Bangalore 560080, India}
\author{Abhishek Dhar}
\email{abhishek.dhar@icts.res.in}
\affiliation{ International centre for theoretical sciences, TIFR, IISC campus, Bangalore 560012}
\author{Keiji Saito}
\email{saitoh@rk.phys.keio.ac.jp}
\affiliation {Department of Physics, Keio University, Yokohama 223-8522, Japan}
\author{ Christian B. Mendl}
\email{mendl@ma.tum.de}
\affiliation{Zentrum Mathematik, TU M\"{u}nchen, Boltzmannstr. 3, 85747 Garching, Germany}  
\author{Herbert Spohn}
\email{spohn@ma.tum.de}
\affiliation{Institute for Advanced Study, Einstein Drive, Princeton NJ 08540, USA} 
\vspace{0.5cm}

\date{\today}

\begin{abstract} 
Recent work has developed  a nonlinear hydrodynamic fluctuation theory for a chain of
 coupled anharmonic oscillators governing the conserved fields, namely  
stretch, momentum, and energy. The linear theory yields two propagating sound modes and one diffusing heat mode. 
In contrast, the nonlinear theory predicts that, at long times, the sound mode correlations satisfy Kardar-Parisi-Zhang (KPZ) scaling, while the heat mode correlations 
satisfies L\'{ev}y-walk scaling. In the present contribution we report on  molecular dynamics simulations of Fermi-Pasta-Ulam chains to compute various 
spatiotemporal
correlation functions and compare them with the predictions of the theory.  We find very good agreement in many cases, but
also some deviations.
\end{abstract}

\maketitle

\section{Introduction}
It is now  general consensus that heat conduction in one-dimensional
(1D) momentum conserving systems is anomalous \cite{LLP03,dhar08}. 
There are various approaches which lead to this conclusion. The first approach is through direct nonequilibrium molecular dynamics 
simulations \cite{LLP97,dhar02,grass02,casati03,mai07}. Consider a system of $N$ particles  connected at the ends to heat baths with a 
small temperature difference $\Delta T$, so that  a steady state heat current $J$ flows across the system. Defining the thermal conductivity
as $\kappa = JN/\Delta T$ one typically finds 
\bea
\kappa \sim N^\alpha
\eea
with $0<\alpha<1$, which means that Fourier's law is not valid. 
A second approach is to use the Green-Kubo formula relating thermal conductivity
to the integral over the equilibrium heat current auto-correlation function.  Simulations and several 
theoretical approaches  \cite{delfini06,narayan,wang04,pereverez03,lukkarinen,BBO,beijeren} find that the correlation function 
has a slow power law decay $\sim 1/t^{1-\alpha}$ and this again results in a divergent conductivity.  
Finally, a number of contributions \cite{onuttom03,zhao06,chen11,cipriani,denisov11,lepri11,dhar13,liu14} have studied the decay of equilibrium energy 
fluctuations or of  heat pulses and find that they are super-diffusive. This is understood through phenomenological models in which the energy 
carriers perform  L\'{e}vy walks \cite{cipriani, denisov11, lepri11,dhar13}.

A significant step towards understanding anomalous heat transport in one-dimension was achieved recently in
\cite{beijeren} and extended to anharmonic chains in \cite{mendl13,spohn13}, 
where a detailed theory of hydrodynamic fluctuations is developed including several analytic results. 
The main strength of this theory lies in very detailed predictions which can be  verified through direct 
simulations of 
microscopic models. Unlike earlier studies which have mainly focused on the thermal conductivity exponent $\alpha$, 
nonlinear fluctuating hydrodynamics predicts the scaling forms of various correlation functions,  including prescriptions to compute the non-universal 
parameters for a given microscopic model.  The hydrodynamic theory is based on several assumptions and hence there is a 
need to check the theory 
through a comparison with results from molecular dynamic simulations. This is the aim of our contribution. In a recent paper \cite{mendl14} 
results are discussed for hard point particle systems either interacting via the so-called shoulder potential or with alternating masses.  Here we consider  
Fermi-Pasta-Ulam (FPU) chains, report on simulation results for equilibrium time correlations in different parameter regimes, and compare with the theory. 

There is a large body of work which addresses the  equilibration in FPU chains \cite{berman,benettin,ruffo}. As in our study they start from random initial data. However they consider non-equilibrium initial conditions at very low energy. In contrast we investigate the decay of time correlations with initial data chosen from a thermal distribution at a moderate temperature. The correlation functions are obtained by performing an average over these 
initial conditions.

Let us first summarize the results of the theory in \cite{spohn13}. Consider $N$ particles with positions and momenta described by the 
variables $\{q(x),p(x)\}$, for $x=1,\ldots,N$,  and moving on a periodic ring of size $L$ such that $q({N+1})=q(1)+L$ and $p({N+1})=p(1)$. Defining the 
``stretch'' variables $r(x)=q(x+1)-q(x)$, the anharmonic chain is described by the following Hamiltonian with nearest neighbor interactions
\bea  \label{ham}
H = \sum_{x=1}^N \epsilon(x), \quad
\epsilon(x) = \f{p^2(x)}{2}+ V[r(x)]~, 
\eea 
where the particles are assumed to have unit mass. From the Hamiltonian equations of motion one conludes that stretch $r(x)$, 
momentum $p(x)$, and energy $\epsilon(x)$ are locally conserved and satisfy the following equations of motion
\begin{align}
\f{\p r(x,t)}{\p t}&=\f{\p p(x,t)}{\p x}, \nn \\
\f{\p p(x,t)}{\p t}&=-\f{\p {P}(x,t) }{\p x},\nn \\
\f{\p e(x,t)}{\p t}&=-\f{\p}{\p x}[ p(x,t) {P}(x,t)]~, \label{eqm}
\end{align} 
where ${P}(x)=-dV(r)/dr|_{x-1}$ is the local force and $\p f/\p x=f(x+1)-f(x)$ denotes the discrete derivative. 
Assume that the system is in a state of thermal equilibrium at zero total average momentum in such a way
 that,  respectively, the average energy 
and average stretch are fixed by the temperature ($T=\beta^{-1}$) and pressure ($P$) of the chain. 
This corresponds to an ensemble defined by the distribution 
\bea\label{measure} 
\mathcal{P}(\{p(x),r(x)\})=\prod_{x=1}^N \frac{e^{-\beta[p_x^2/2+V(r_x)+P r_x]}}{Z_x}~, 
\quad Z_x=\int_{-\infty}^\infty dp \int_{-\infty}^\infty dr e^{-\beta [ p^2/2+V(r)+P r]}~. 
\eea 

Now consider small fluctuations of the conserved quantities about their equilibrium values,
 $u_1(x,t)=r(x,t)- \la r \ra_{eq}$, $u_2(x,t) = p(x,t)$ and  $u_3(x,t)=\epsilon(x,t) -\la \epsilon \ra_{eq}$. 
The fluctuating hydrodynamic equations for the field $\vec{u}=(u_1,u_2,u_3)$ are now 
written by expanding the conserved currents in Eq.~(\ref{eqm}) to second order in the nonlinearity and then adding dissipation and noise terms to ensure 
thermal equilibration. 
Thereby one arrives at the noisy hydrodynamic equations
\bea
\p_t u_\alpha= -\p_x  \left[ A_{\alpha \beta} u_\beta + H^\alpha_{\beta \gamma}  u_\beta u_\gamma - \p_x \widetilde{D}_{\alpha \beta} u_\beta + 
\widetilde{B}_{\alpha \beta} \xi_\beta \right]~.~~~~\label{EOM}
\eea
The noise and dissipation matrices, $\widetilde{B},\widetilde{D}$, are related by the
fluctuation-dissipation relation $\widetilde D C + C \widetilde{D} = \widetilde{B} \widetilde{B}^T$, where the matrix $C$ corresponds to equilibrium correlations and has elements  $C_{\alpha \beta}(x)= \la u_\alpha(x,0) u_\beta(0,0)\ra$. 

We switch to normal modes of the linearized equations through the 
transformation $(\phi_{-1},\phi_0,\phi_1)=\vec{\phi} = R \vec{u}$, where the matrix $R$ acts only on the component index and diagonalizes $A$, \ie~ 
$R A R^{-1}={\rm diag}(-c,0,c)$. The diagonal form implies that there are two 
sound modes, $\phi_\pm$, traveling at speed $c$ in opposite directions and one stationary but decaying heat mode, $\phi_0$.
The quantities of interest are the equilibrium spatiotemporal 
 correlation functions $C_{s s'}(x,t)=\la \phi_s(x,t) \phi_{s'}(0,0)\ra$, where $s,s'=-,0,+$. Because the modes separate linearly in time, one argues that they decouple into three single component equations. These have the structure of the noisy Burgers equation, for which the exact scaling function, denoted by $f_\mathrm{KPZ}$, is available. This works well for the sound peaks. But for the heat peak the self-coupling coefficient vanishes whatever the interaction potential. Thus one has to study the sub-leading corrections, which at present can be done only within mode-coupling approximation, resulting in the symmetric L\'{e}vy-walk distribution. 
While this is an approximation, it seems to be very accurate. For the generic case of non-zero pressure, \ie~$P\neq0$,
 which corresponds either to asymmetric inter-particle potentials or to an externally applied stress, the prediction
 for the left moving, resp. right moving, sound peaks and the heat mode are  
\begin{align} 
  C_{\mp\mp}(x,t)&=   \f{1}{(\lambda_s t)^{2/3}}~ f_{\mathrm{KPZ}} \left[~ \f{(x \pm ct)}{(\lambda_s t)^{2/3}}~\right]~, \label{eqscalS}\\
  C_{00}(x,t) &=  \f{1}{(\lambda_e t)^{3/5}} ~f^{5/3}_{\mathrm{LW}}\left[~ \f{x}{(\lambda_e t)^{3/5}}~\right]~.\label{eqscalE} 
\end{align}
$f_{\mathrm{KPZ}}(x)$ is the KPZ scaling function discussed in \cite{spohn13,prahoferspohn2004}, and tabulated in \cite{prahofertable}.
$f_{\mathrm{LW}}^\nu(x)$ is the Fourier transform of the L\'{e}vy characteristic function $e^{-|k|^\nu}$.  
For an even potential at $P = 0$, all self-coupling coefficients vanish and, within mode-coupling approximation,
one obtains
\begin{align} 
 C_{\mp\mp}(x,t)&= \frac{1}{{(\lambda^0_s t)}^{1/2}} f_{\mathrm{G}} \left[~ \f{(x \pm ct)}{(\lambda_s^0 t)^{1/2}}~\right]~,\label{eqscalSP0}\\
  C_{00}(x,t) &=  \f{1}{(\lambda_e^0 t)^{2/3}} ~f^{3/2}_{\mathrm{LW}}\left[~ \f{x}{(\lambda_e^0 t)^{2/3}}~\right]~,\label{eqscalEP0}
\end{align} 
 where $f_\mathrm{G}(x)$ is the unit Gaussian with zero mean. In a recent contribution \cite{JKO}, a model with the same signatures is studied and their exact result agrees with the mode-coupling predictions (\ref{eqscalSP0}), (\ref{eqscalEP0}).

For the non-zero pressure case, the scaling coefficients $\lambda_s$ and $\lambda_e$ for the sound and heat mode respectively are given by
\begin{align} 
\lambda_s &= 2 \sqrt{2} |G^1_{11}|~,\nn \\ 
\lambda_e &={\lambda_s}^{-2/3} C^{-1/3} {(G^0_{1 1})}^2 a_e~, \label{lameq}
\end{align} 
where $a_e=2 \sqrt{3} ~\Gamma(1/3) \int^{\infty}_{-\infty}~dx {f_{KPZ}(x)}^2 = 3.167...$ is a model-independent numerical constant, and the 
matrices $G^\alpha$ are related to the nonlinear coupling matrices $H^\alpha$ through the normal mode transformation defined by $R$
(see Appendix for details).\\
For the case of an even potential at zero pressure, the sound peaks are diffusive. The coefficient  $\lambda_s^0$ is a transport coefficient which in principle is determined through a Green-Kubo formula. Thus an explicit formula
is unlikely and $\lambda_s^0$ remains undetermined by the theory. In contrast, for the generic case the leading coefficients are obtained from static averages. The heat mode couples to the sound modes and
its exact scaling coefficient is 
\begin{align}
\lambda_e^0 = (\lambda_s^0)^{-1/2} c^{-1/2} {(G^0_{11})}^2 {(4\pi)}^2 a_e^0,~\label{lambdaeP0}
\end{align}
where $a_e^0= 4 \int^\infty_{0} dt~t^{-1/2} cos(t) \int^\infty_{-\infty} dx~{f_G(x)}^2 = \sqrt{2}$. From a simulation 
of the microscopic dynamics one obtains $\lambda_s^0$, and from there one calculates $\lambda_e^0$ using the above formula.
In the following section, we discuss the results of our molecular dynamics simulations in  computing the correlation functions
and compare them with the scaling predictions.

\section{Molecular Dynamics Simulations}

To verify the predictions from hydrodynamics, we consider the FPU 
$\alpha$-$\beta$ model described by the following  inter-particle potential
\bea
V(r) =  k_2 \f{r^2}{2} +  k_3  \f{r^3}{3} + k_4 \f{r^4}{4} ~.
\eea
The set of variables $\{r(x),p(x)\}$, $x=1,2,\ldots,N$,  
 are evolved  according to the equations of motion
\bea
\dot{r}(x)=p(x+1)-p(x),~~\dot{p}(x)=V'(r(x))-V'(r(x-1))~,
\eea
with  
initial conditions chosen from the distribution given by Eq.~(\ref{measure}).
For the product measure it is easy to generate the initial distribution directly and one does not need to dynamically equilibrate the system. 
The integrations have been done using both the velocity-Verlet algorithm \cite{AT87} and also through the fourth order Runge-Kutta algorithm 
and we do not find any 
significant difference. 
The full set of two-point correlation functions were obtained by averaging over around $10^6-10^7$ initial conditions. 
Here we present results for four different parameter sets.

\begin{figure}
\vspace{1cm}
  \includegraphics[width=3.2in]{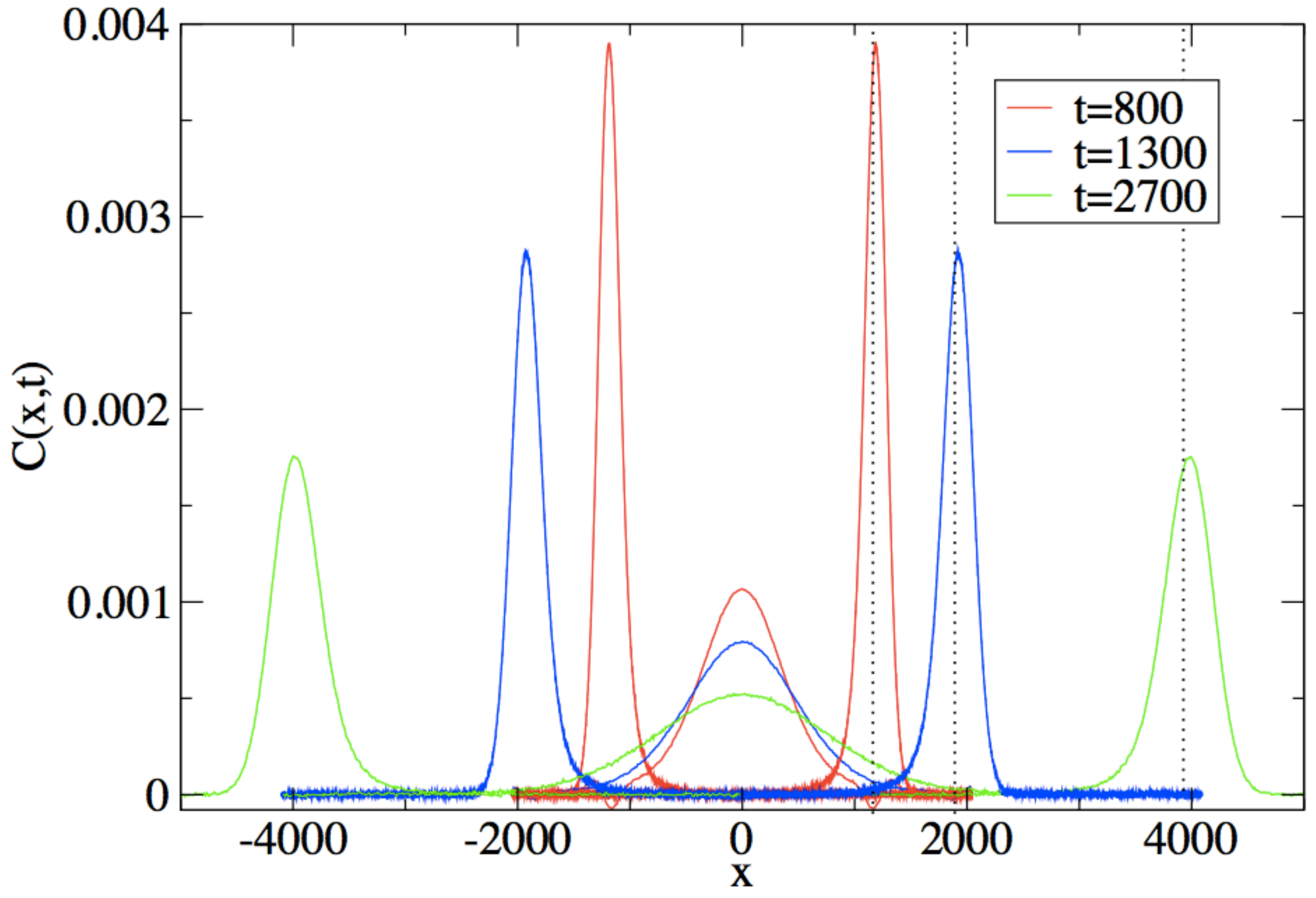}
\caption{ Set I: The parameters of the simulation are $k_2=1,~ k_3=2,~ k_4=1,~T=0.5$, $P=1$ and system size $N=8192$.
Correlation functions for the heat mode and the two sound modes  at three different times. At the latest time we see that the heat and 
sound modes are well separated.  The speed of sound is $c=1.45468$. }
\label{S1All}
\end{figure}

\begin{figure}
\includegraphics[width=3.2in]{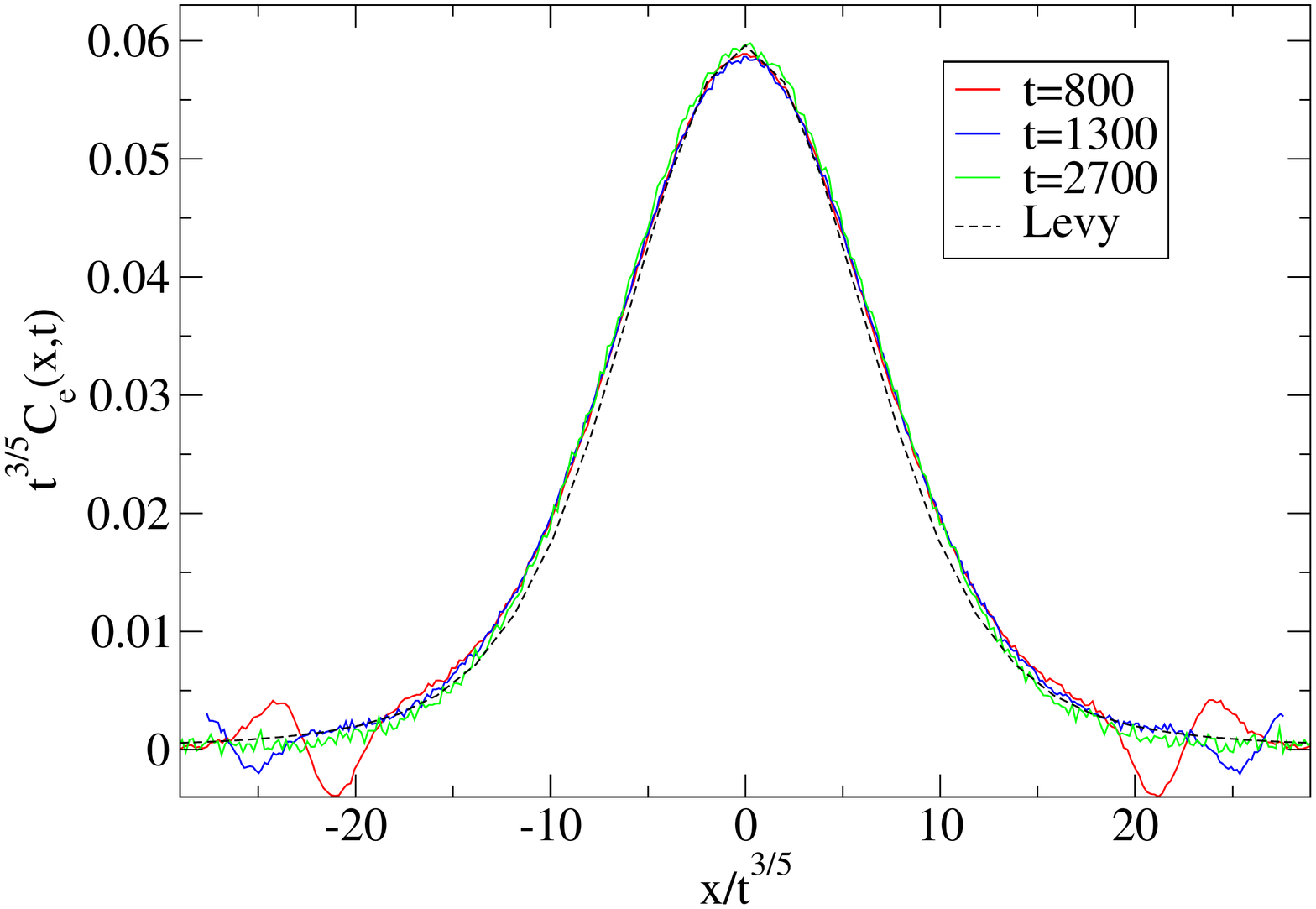} 
\includegraphics[width=3.2in]{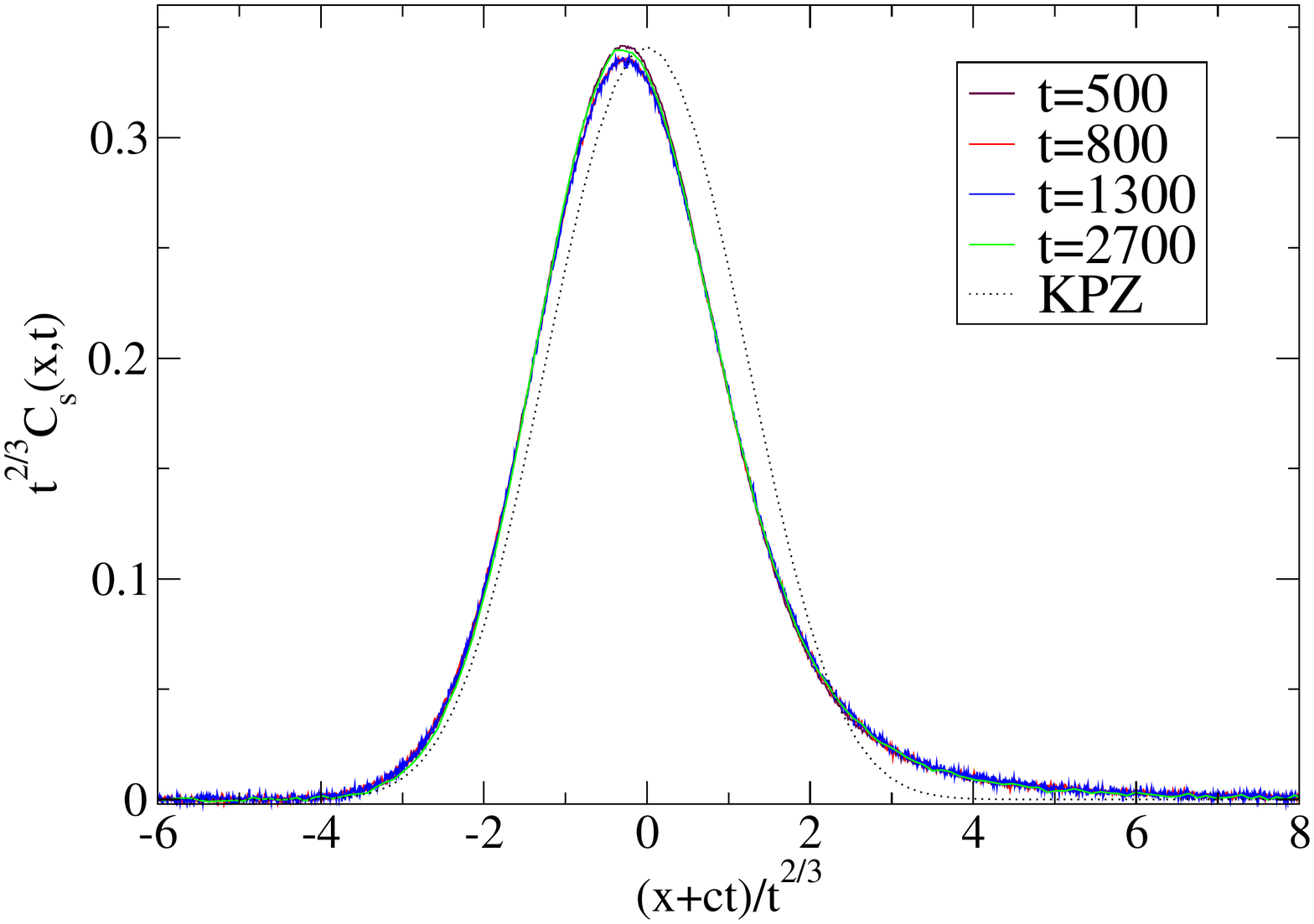}
\caption{ Set I: Same parameters as in Fig.(\ref{S1All}).
Scaled plots of heat mode and left moving sound mode correlations, 
at different times, using a L\'{e}vy-type-scaling for the heat mode and KPZ-type scaling for the sound mode.  We see here that the collapse of different time data is very good. 
The fit to the L\'{e}vy-stable distribution with $\lambda_e=13.8$ is quite good, while the fit to the KPZ scaling function with 
$\lambda_s=2.05$ is not convincing. }
\label{S1scalA}
\end{figure}

{\bf Set I}:~$k_2=1.0,k_3=2.0,k_4=1.0,T=0.5,P=1.0$. These are the set of parameters used in \cite{mendl13} for the numerical  solutions of the mode-coupling equations. In Fig.~(\ref{S1All}) we show the heat mode correlation $C_{00}$ and the sound mode correlations
$C_{--},C_{++}$ at three different times. The speed of sound is $c=1.45468...$ The dotted vertical lines in the figure indicate the distances 
$\ell=ct$. The sound peaks are at their anticipated positions. In Fig.~(\ref{S1scalA}) we show the heat mode and the  left moving sound mode after scaling according 
to the predictions in Eqs.~(\ref{eqscalS},\ref{eqscalE}). One can see that the scaling is very good. For comparison we have also plotted a 
L\'{e}vy-stable distribution and the KPZ scaling function \cite{prahofertable},
and find that the agreement is good for the heat mode but not so good for the sound mode. 
One observes a still 
significant asymmetry in the sound mode correlations, contrary to what one would expect from the symmetric KPZ function. 

From our numerical fits shown in Fig.~(\ref{S1scalA}) we obtain the estimates 
$\lambda_s=2.05$ and $\lambda_e=13.8$. 
The theoretical values based on  Eq.~(\ref{lameq})  are  $\lambda_s=0.675$ and $\lambda_e=1.97$ (see Appendix), which thus deviates significantly from the numerical estimates obtained from the simulations.
The disagreement could mean that, for this choice of parameters, we are still not in the asymptotic hydrodynamic regime. 
We expect that the scaling will improve if the heat and sound mode  
are more strongly decoupled. To check this, we simulated a set of parameters where the sound speed is higher and the separation between the sound and heat modes is 
more pronounced. We now discuss this case. 

\begin{figure}
\includegraphics[width=3.2in]{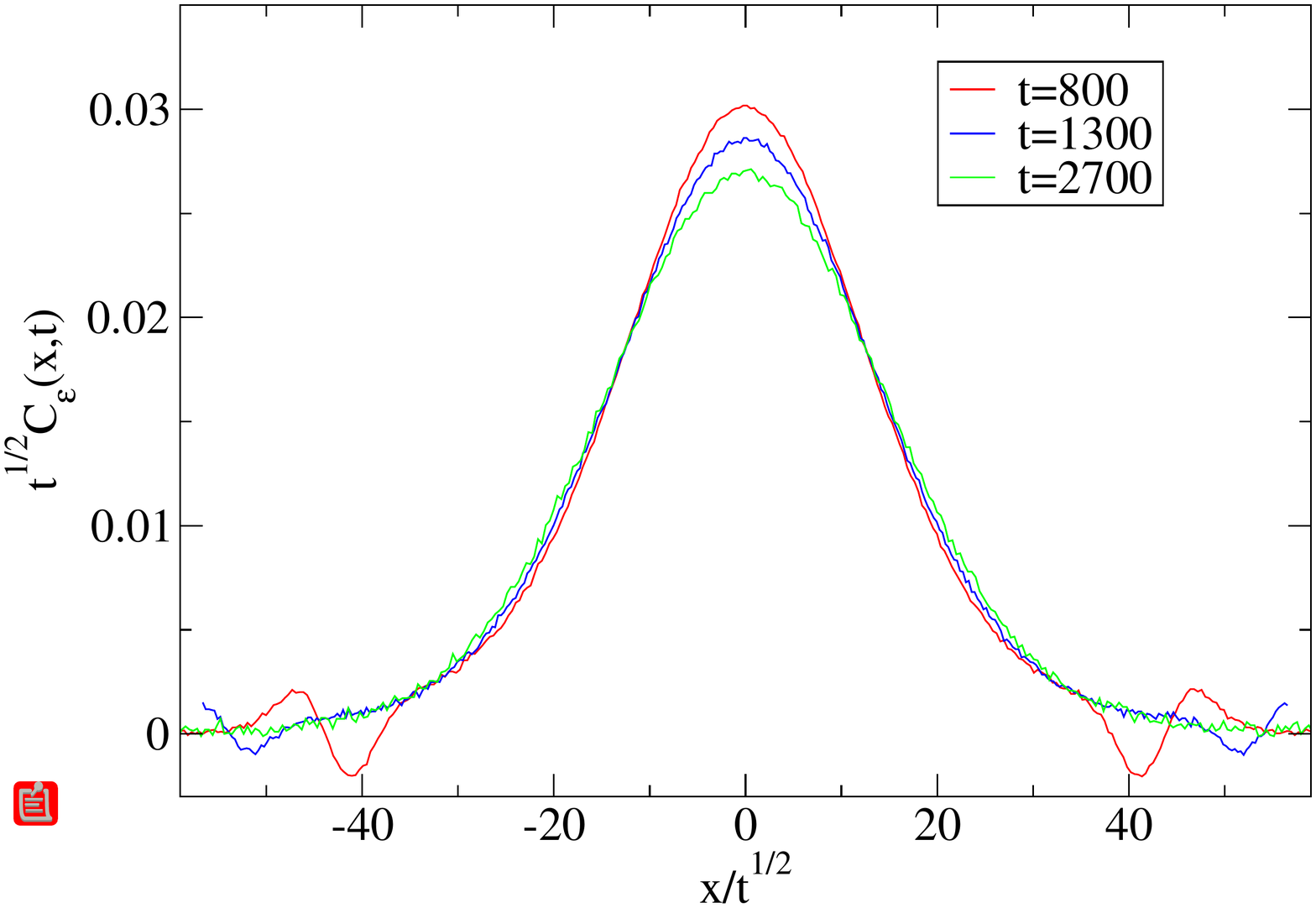} 
\includegraphics[width=3.2in]{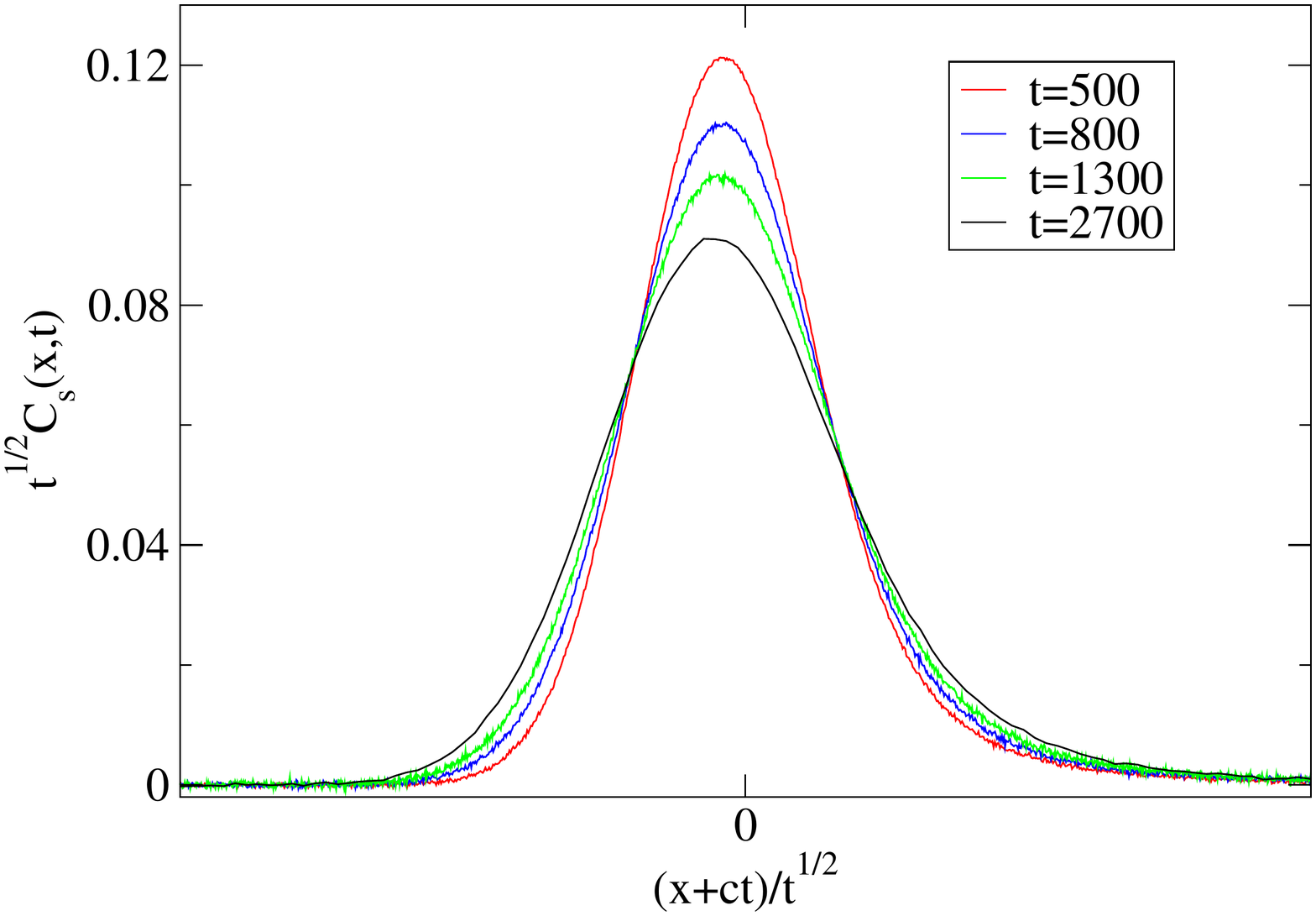}
\caption{Set I: Same parameters as in Fig.(\ref{S1All}). Scaled plots of heat mode and right moving sound mode correlations, 
at different times, using a diffusive scaling ansatz.  We see here that the collapse of different time data is not very good and so clearly the modes are not diffusive.
}
\label{S1scalB}
\end{figure}

\begin{figure}
\vspace{1cm}
\includegraphics[width=3.2in]{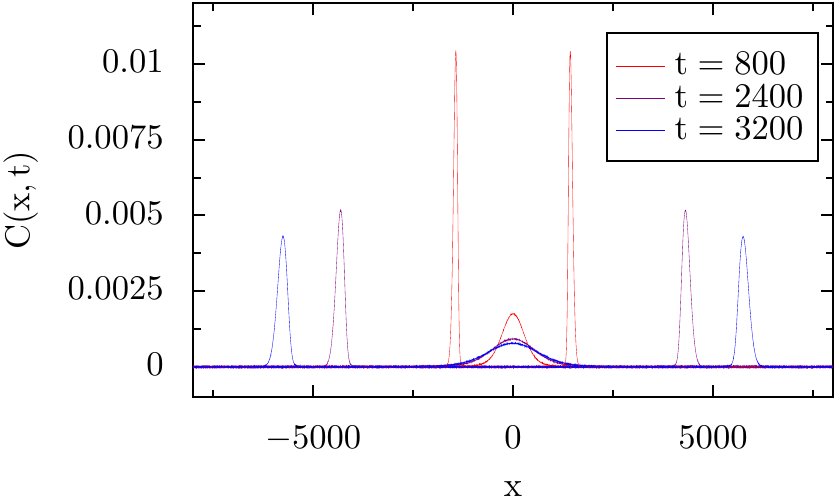}
\caption{ Set II: Heat and sound mode correlations at three different times for the parameter set as in 
Fig.~(\ref{S1All}) but with $T=5.0$ and system size $16384$. The speed of sound in this case was $c=1.80293$. In this case we see that the separation
of the heat and sound modes is faster and more pronounced than for the parameter set of Fig.~(\ref{S1All}). }
\label{S2All}
\end{figure} 

\begin{figure}
\includegraphics[width=3.4in]{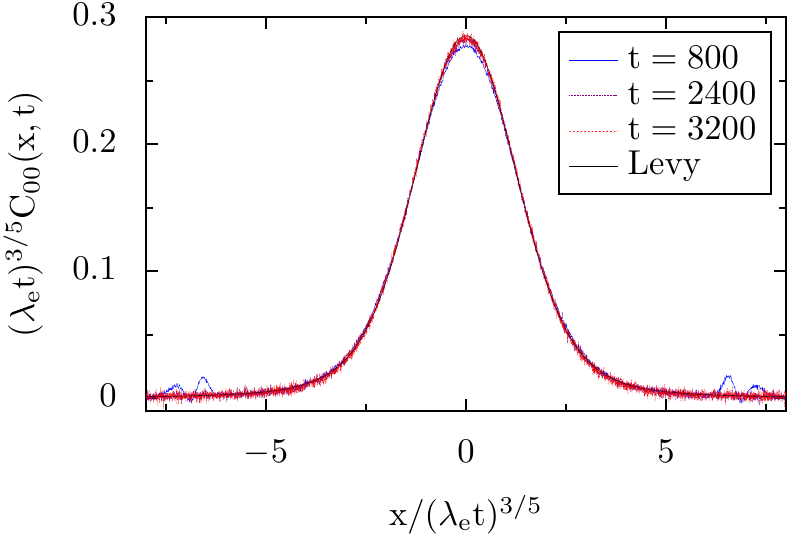} 
\includegraphics[width=3.4in]{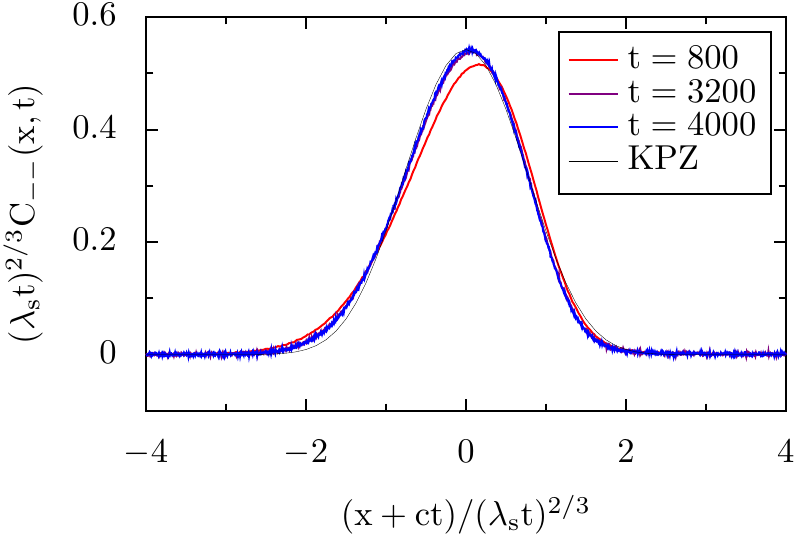}
\caption{Set II:  Same parameters as in Fig.(\ref{S2All}).
Scaled plots of heat mode and left moving sound mode correlations
at different times, using a L\'{e}vy-type-scaling for the heat mode and KPZ-type scaling for the sound mode. 
We see here that the collapse of different time data is very good. 
Again we find a very good fit to the L\'{e}vy-stable distribution with $\lambda_e=5.86$ while the  fit to the KPZ scaling function, with $\lambda_s=0.46$, is  not yet perfect.}
\label{S2scalA}
\end{figure}

\begin{figure}
\vspace{1cm}
\includegraphics[width=3.2in]{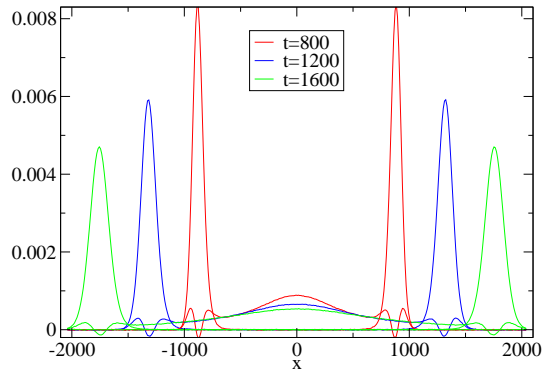}
\caption{Set III: Low temperature case - The parameters of the Hamiltonian are $k_2=1,~ k_3=-1,~ k_4=1,~T=0.1$, $P=0.07776$ and $N=4096$. In this plot we show the heat mode correlation and the two sound mode correlations at three different times. In this case the separation between heat and sound modes is less pronounced.
} \label{LTAll}
\end{figure}

\begin{figure}
\includegraphics[width=3.2in]{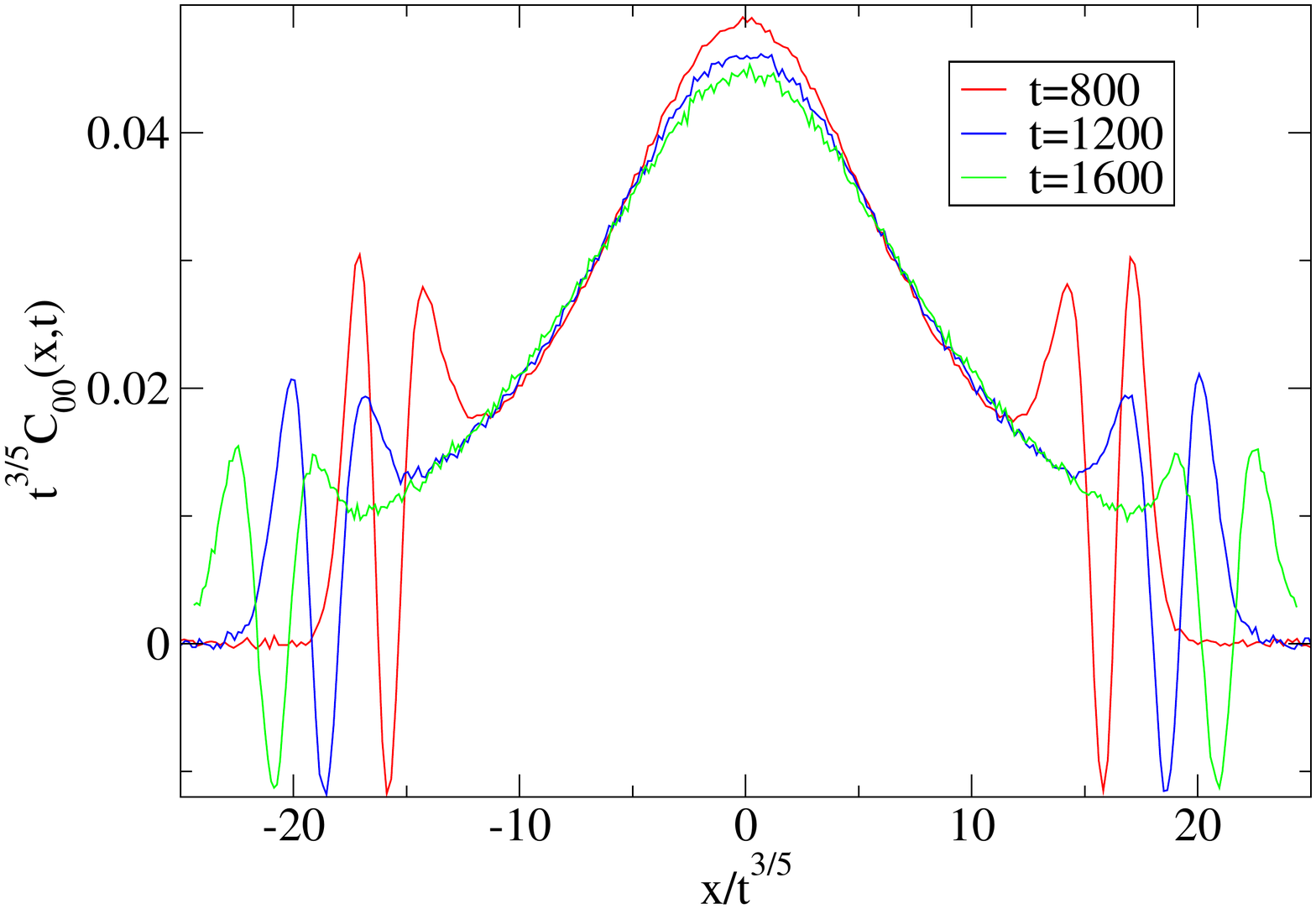} 
\includegraphics[width=3.2in]{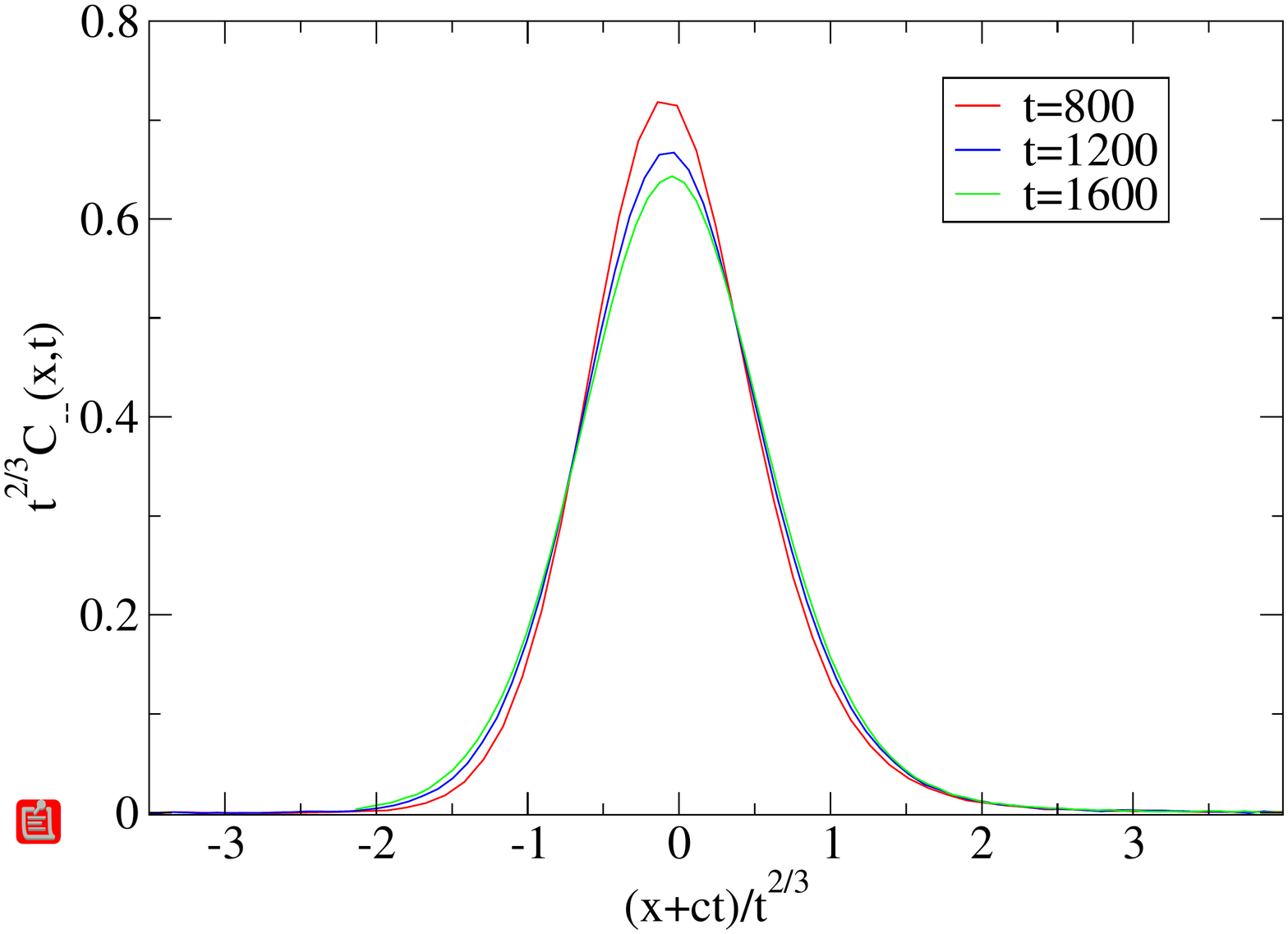}
\caption{Set III: Same set of parameters as in Fig.~(\ref{LTAll}). 
Scaled plots of heat mode and left moving sound mode correlations, 
at different times, using a L\'{e}vy-type-scaling for the heat mode and KPZ-type scaling for the sound mode. We see here that the collapse of different time data for the heat mode is reasonably good. }
\label{LTscalA}
\end{figure} 

\begin{figure}
\includegraphics[width=6.4in,height=2.2in]{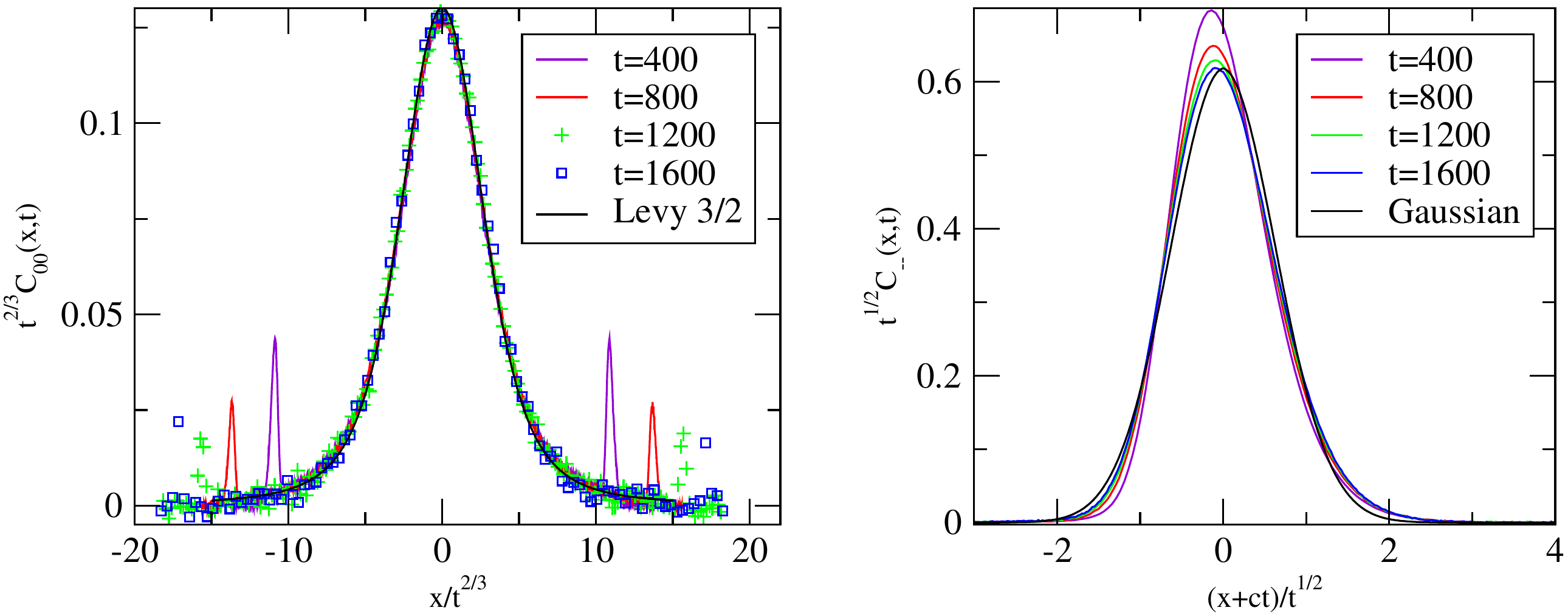}
\caption{Set IV: Even potential, zero pressure case - The parameters of the Hamiltonian are $k_2=1, k_3=0, k_4=1$,  $P=0$, $T=1$ and $N=8192$. 
The scaling used  here corresponds to Eqs.~(\ref{eqscalSP0},\ref{eqscalEP0}), with $\lambda_s^0=0.416$ and $\lambda_e^0=3.18$.} 
\label{P0scalA}
\end{figure}

\begin{figure}
\includegraphics[width=6.4in,height=2.2in]{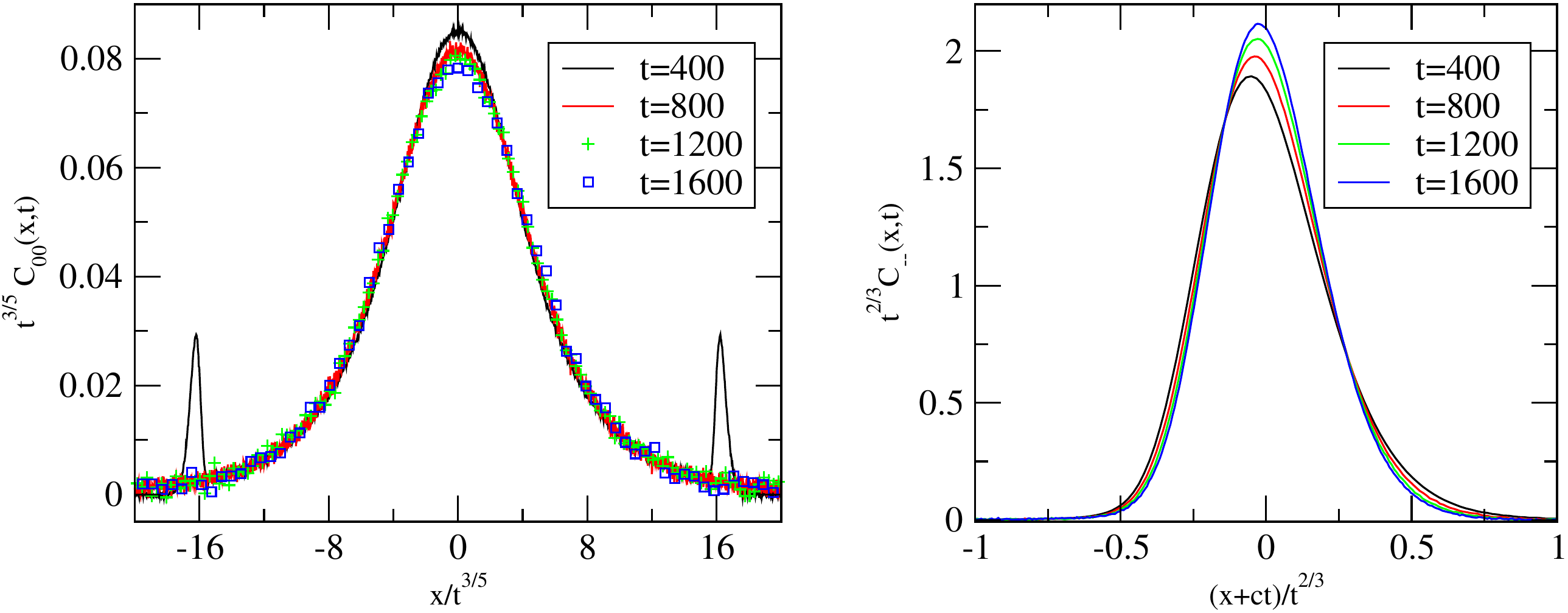}
\caption{Set IV: Parameters same as in Fig.~(\ref{P0scalA}).
The scaling used  here corresponds to Eqs.~(\ref{eqscalS},\ref{eqscalE}). 
We see that the collapse is not as good as in Fig.~(\ref{P0scalA}).}
\label{P0scalB}
\end{figure} 

\begin{figure}
\includegraphics[width=6.4in,height=2.2in]{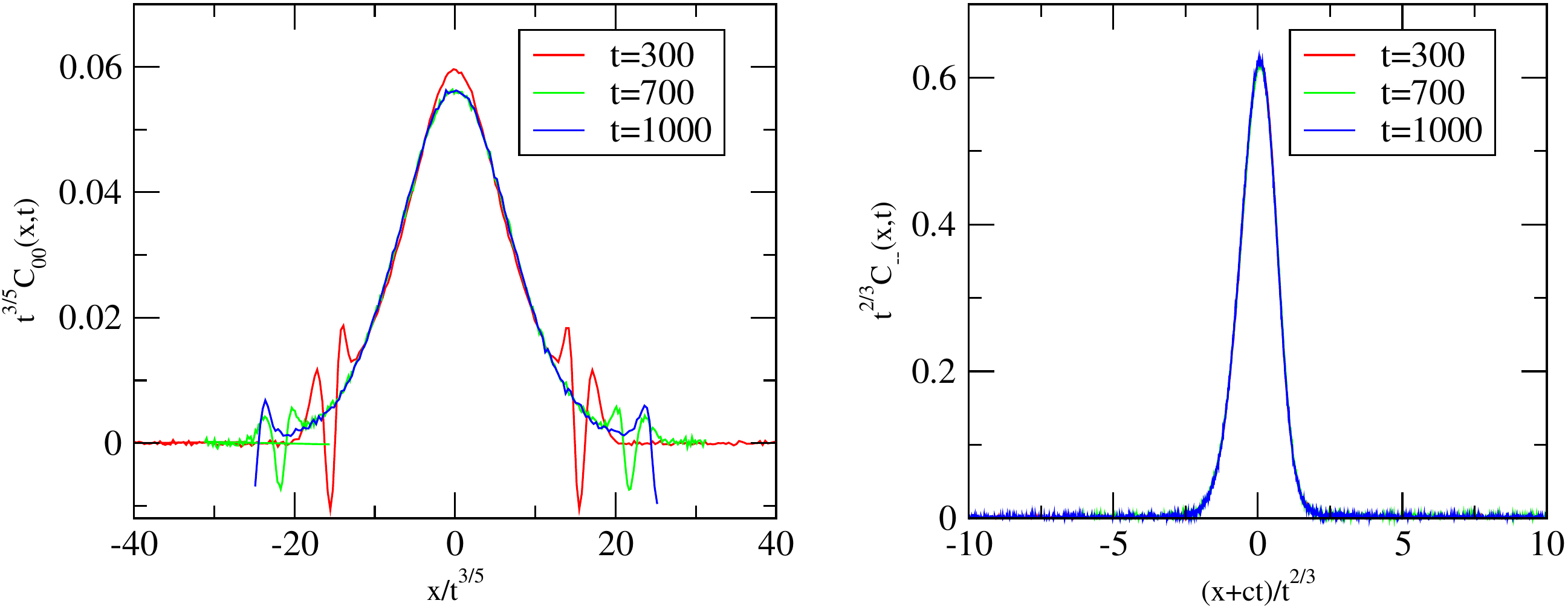}
\caption{Set V: Even potential, finite pressure case - The parameters of the Hamiltonian are $k_2=1, k_3=0, k_4=1$, $P=1$, $T=1$ and $N=3200$. 
In this case  the scaling corresponding to  Eqs.~(\ref{eqscalS},\ref{eqscalE}) is used in this figure and we have checked  that the scaling  in Eqs.~(\ref{eqscalSP0},\ref{eqscalEP0}) does not work as well.} 
\label{evenP0}
\end{figure}

{\bf Set II}: $k_2=1.0,k_3=2.0,k_4=1.0,T=5.0,P=1.0$. This choice of parameters 
gives $c=1.80293$ and we see in Fig.~(\ref{S2All}) there is much better separation of the heat and sound modes. We again find an excellent collapse of the heat mode and the sound mode data with the expecteded scalings Fig.~(\ref{S2scalA}) . The heat mode  fits very well to the L\'{e}vy-scaling function. 
However the sound-mode scaling function  still shows significant asymmetry and is different from the KPZ function.  
The theoretical obtained values of $\lambda_s=0.396$  and $\lambda_e=5.89$  
are now close to the numerically estimated values $\lambda_s=0.46$ and $\lambda_e=5.86$.

{\bf Set III}: $k_2=1.0,k_3=-1.0,k_4=1.0,T=0.1,P=0.07776$. Our third choice of the parameter set is motivated by recent nonequilibrium simulations 
\cite{zhao12,zhao13} which find that the thermal conductivity $\kappa$ at  low temperatures seems to converge to 
 a size-independent value, contradicting the expectation that heat conduction is anomalous and $\kappa$ should diverge with system size at all temperatures. 
 It has been suggested that this 
could be a finite size effect \cite{das14,wang13,savin14}, but this has not been established convincingly yet. 
Here we want to explore if the equal-time  correlations show any signatures of  diffusive heat transport and if they provide any additional insight regarding the strong finite size effects seen in the nonequilibrium studies. 
The temperature chosen is $T=0.1$, which for the FPU potential parameters above correspond to the regime at which normal conduction
has been proposed. 

The speed of sound is calculated to be $c=1.09352$, which matches with the numerical data, as seen in Fig.~(\ref{LTscalA}).
The heat mode seems to follow the predicted anomalous scaling reasonably well (though the convergence is, as 
expected, slower than in the high-temeprature case). We have checked that the same data, when scaled as $t^{1/2}C_{00}(x/t^{1/2})$ for different times,
shows no indication of convergence. Thus we find no evidence for normal heat diffusion at low temperatures.
The sound mode agrees quite well with the KPZ-type scaling observed for
higher temperatures, though the shape of the correlation function remains asymmetric as in the high-temperature case. 

It will be noted that the heat mode shows two peaks near the edges which do not follow the L\'{e}vy scaling - these peaks arise from
interaction with the sound modes, indicating that there is still some overlap between the two modes near the edges.
The sound mode, on the other hand, is found to be undistorted, which is consistent with the prediction from \cite{spohn13} that at long times
the mode-coupling equations for the sound modes becomes independent of the heat mode, but not vice versa. the same effect can be seen in sets I and II,
but are less pronounced as the heat and sounds separate more quickly at higher temperatures.

{\bf Set IV}: $k_2=1.0,k_3=0.0,k_4=1.0,T=1.0,P=0.0$. This is the special case of an even potential at zero pressure for which 
the prediction from the theory is a diffusive sound mode, while the heat mode is L\'{e}vy but with a different exponent. 
The predicted scalings are given in Eqs.~(\ref{eqscalSP0},\ref{eqscalEP0}). The speed of sound
for this case is $c=1.46189$. We see from Fig.~(\ref{P0scalA}) that the proposed scaling
leads to an excellent collapse of the heat mode at different times. The sound mode, with diffusive scaling, shows a strong convergence but not yet a collapse.
Fig. \ref{P0scalB} shows the same data but scaled according to the predictions in the non-zero pressure case. It is clear that the data are non-convergent 
with this scaling. 

The sound mode is predicted by the theory to be Gaussian, Eq.~(\ref{eqscalSP0}), but as seen from Fig.~(\ref{P0scalA}), the fit
to the Gaussian form is poor. From the data we estimate that $\lambda_s^0=0.416$, and upon using Eq.~(\ref{lambdaeP0}), we find 
$\lambda_e^0=1.17$, whereas the numerically obtained value is $3.18$.

{\bf Set V}: $k_2=1.0,k_3=0.0,k_4=1.0,T=1.0,P=1.0$ Parameters are identical to the above set, except that the pressure is non-zero. Since the potential 
is even,
the pressure arises from externally applied stress to the system. The speed of sound is $c=1.59143$. We find in Fig. \ref{evenP0} that the 
correlations satisfy the same scaling
as a generic asymmetric potentials with non-zero pressure (sets I, II, III). 
This confirms that the universality class is determined by the asymmetry of $V(r) +Pr$ and not of $V(r)$ by itself.

\section{Discussion}
We have performed numerical simulations of FPU chains to test the predictions of nonlinear fluctuating hydrodynamics
in one dimension \cite{spohn13}. The theory predicts the existence of a zero velocity heat mode and two mirror image
moving sound modes, and provides their asymptotic scaling forms. 
We have tested the theory for various parameter regimes, including high and low temperatures, and zero and non-zero pressure regimes. For non-zero pressure, 
we find that
the heat mode scales according to the L\'{e}vy-5/3 distribution, as predicted by the theory, both at high and low temperatures. This implies that 
for one-dimensional heat transport (in momentum-conserving systems) the scaling is generically anomalous. There are no signatures  of a non-equilibrium 
phase transition (or crossover) from  anomalous to normal conduction.  For the case of an even potential at  zero pressure, the heat mode scales according to
the L\'{e}vy-3/2 distribution as predicted, thus confirming the existence of a second universality class for heat transport in one-dimensional
momentum-conserving systems.

 For non-zero pressure the sound mode spatio-temporally 
scales with the same exponents as the KPZ function, but the shape of the modes is observed not to have collapsed to the KPZ function. This could be
because the simulation times are not in the asymptotic regime for the sound modes, which would be consistent with the slowly decaying correction terms 
to the scaling forms of the sound mode as discussed in \cite{spohn13}. Thus the prediction that the 
sound mode correlations scale according to the KPZ function is not conclusively verified.  The case of an even potential at  zero pressure is very similar, with the sound mode satisfying diffusive scaling, but the limit of the Gaussian scaling shape is not reached in our simulations.

 Although the L\'{e}vy stable distribution fits the heat modes very well, we find that at low temperatures the theoretically predicted values for the 
scaling coefficients $\lambda_s$ and $\lambda_e$ do not well match the numerical values. This is consistent with the numerical study in \cite{mendl14}, where the authors find that for certain hard-point potentials 
the scaling shape has an excellent match, but the scaling coefficients are still drifting and one might hope them to converge to the predicted values at 
larger times. However at high temperatures where the modes are well-separated, the theoretical $\lambda_e$ matches very well with the numerical data,
and the theoretical $\lambda_s$ is not far off from the numerically obtained value.  

An open and important question is to tie up the picture obtained from the correlation dynamics in equilibrium studies with the nonequilibrium properties of FPU chains. The studies here confirm that heat conduction
in one-dimensional chains is anomalous. We do not see any signatures of  
the apparent diffusive behavior (possibly related to finite size-effects) observed in nonequibrium studies at low temperatures \cite{zhao12,das14,wang13}. A clear microscopic understanding of the puzzling strong finite size-effects is
lacking.

\section{Acknowledgements} A.D thanks DST for support through the
Swarnajayanti fellowship. K.S was supported by MEXT (25103003) and JSPS (90312983). We thank Manas Kulkarni for  useful discussions. 

\pagebreak

\section{Appendix} 

The matrix $R$, which diagonalizes the matrix $A$,  is given by
\begin{equation}
 R = \sqrt{\frac{2\beta}{c^2}}\left(\begin{array}{ccc}
 \partial_l p & -c & \partial_e p \\
 \widetilde{\kappa} p &  0 & \widetilde {\kappa} \\
\partial_l p & c &  \partial_e p \end{array} \right),
\end{equation}
where the columns, including the normalization factor, provide the right eigenvectors $V_\alpha$, $\alpha=-1,0,1$, of the $A$ matrix.

The Hessian tensor $H$ encodes the quadratic corrections to the couplings between the original hydrodynamic variables.
$H^\alpha_{\beta \gamma}$ represents the
coupling of the field component $\alpha$ to the field components $\beta$, $\gamma$. The tensor can be represented through 
three $3\times3$ matrices, one for each value of $\alpha$,
\begin{center}
  {$ H^{u_1}=0$, ~$ H^u = \left( \begin{array}{ccc}
 \partial^2_l p & 0 & \partial_l \partial_e p \\
0 &  -\partial_e p & 0 \\
\partial_l \partial_e p & 0 &  \partial^2_e p \end{array} \right)$, ~$ H^e = \left( \begin{array}{ccc}
 0 & \partial_l p & 0 \\
 \partial_l p &  0 & \partial_e p \\
0 & \partial_e p & 0 \end{array} \right)$.} 
\end{center}

After transforming to the normal modes $\vec {\phi}$, the nonlinear hydrodynamic equations become  
\begin{equation}
 \p_t \phi_\alpha= -\p_x  \left[ c_\alpha \phi_\alpha + \langle \vec {\phi}. G^\alpha \vec{\phi} \rangle  
 - \p_x {(D \phi)}_\alpha + {(B \xi)}_\alpha \right]~. \nn ~~~~
\end{equation}

The term in angular brackets is the inner product of $G^\alpha$ with respect to
$\vec{\phi}$. Also, $D=R \widetilde{D} R^{-1}$ and $B=R \widetilde {B}$ satisfy the fluctuation-dissipation
relation $B B^T=2D$. The vector $\vec {c} = (-c,0,c)$.

The tensor $G$ represents the coupling between the normal modes and is given by 
$G^\alpha=\frac{1}{2}\sum_{\alpha^\prime=1}^3 R_{\alpha \alpha^\prime} (R^{-1})^T H^{\alpha^\prime} R^{-1}$. The the elements of
$G^\alpha$ can be represented through cumulants of $V$, $r$ with respect to the single site distribution up to order three,  see \cite{spohn13}.\\

The values of $R$ and $G^0$ and $G^1$ are given below. The elements of $G^{-1}$ are a rearrangement of 
the elements of $G^{1}$ as follows from $G^{-1}_{-\alpha~-\beta}=-G^{1}_{\alpha \beta}$,~
$G^{-1}_{-1 0}=G^{-1}_{0 1}$ and $G^{-1}_{\alpha \beta}=G^{1}_{ \beta \alpha}$ \cite{spohn13}.
The long time behavior is dominated by the diagonal $G$ entries. The off-diagonal entries are irrelevant. 
$G^s_{ss}$ are the self-couplings. Note that $G^0_{00} = 0$,
as claimed before. Also for the even potential at zero pressure case the only leading terms are $G^0_{ss}$, $s = \pm1$.
There is considerable variation in the diagonal matrix elements.
\\

{\bf Set I}
\begin{center}
$ R = \left( \begin{array}{ccc}
-0.7935 & -1. & 0.66118 \\
1.89594 & 0.0 & 1.89594\\
-0.7935 & 1. & 0.66118 \end{array} \right)$, \quad~$ G^0 = \left( \begin{array}{ccc}
 -0.689497 & 0.0 & 0.0 \\
 0.0&  0.0 & 0.0 \\
0.0 & 0.0 &  0.689497 \end{array} \right)$, ~$ G^1 = \left( \begin{array}{ccc}
  -0.24236 & -0.075565 & .238543\\
 -0.075565 & -0.0669417  & -0.075565 \\
.238543 & -0.075565 & 0.238543  \end{array} \right).$
\end{center}

\vspace{1cm}

{\bf {Set II}}
\begin{center}
 
$ R = \left( \begin{array}{ccc}
-0.547157 & -0.316228 & 0.0229798\\
 0.229483 & 0.0 & 0.229483\\
  -0.547157 & 0.316228 & 0.0229798 \end{array} \right)$,\quad~$ G^0 = \left( \begin{array}{ccc}
 - 1.03436& 0.0 & 0.0 \\
   0.0 &  0.0  & 0.0 \\
   0.0 &  0.0  & 1.03436 \end{array} \right)$, ~$ G^1 = \left( \begin{array}{ccc}
 -.0671336 & .240399 & .140022 \\
 .240399 &  -.152971 & .240399 \\
.140022 & .240399 &  .140022 \end{array} \right).$
\end{center}
\vspace{1cm}

{\bf {Set III}}
\begin{center}

$ R = \left( \begin{array}{ccc}
 -2.3376 & -2.23607 & 1.38344\\
 0.793106 & 0.0 & 10.1994;\\
  -2.3376 & 2.23607 & 1.38344 \end{array} \right)$,\quad ~$ G^0 = \left( \begin{array}{ccc}
 -0.55766 & 0.0 & 0.0 \\
 0.0 &  0.0 & 0.0 \\
 0.0 &  0.0 & 0.55766 \end{array} \right)$, ~$ G^1 = \left( \begin{array}{ccc}
 -0.0721968 & 0.0206018 & 0.0790847 \\
 0.0206018 & -0.0353259 &  0.0206018\\
0.0790847 &  0.0206018 & 0.0790847\end{array} \right).$
\end{center}
\vspace{1cm}
{\bf {Set IV}}

\begin{center}
$ R = \left( \begin{array}{ccc}
 -1.03371 & -0.707107 & 0.0\\
  0.0 & 0.0 &1.09893 \\
-1.03371 & 0.707107  & 0.0  \end{array} \right)$,\quad~$G^0 = \left( \begin{array}{ccc}
 -0.803254 & 0.0 & 0.0\\
  0.0 & 0.0 & 0.0 \\
  0.0 & 0.0  & 0.803254  \end{array} \right)$, ~$ G^1 = \left( \begin{array}{ccc}
 0.0 & 0.133622 & 0.0\\
 0.133622 & 0.0 & .133622 \\
 0.0 & .133622 &  0.0 \end{array} \right)$.
\end{center}
\vspace{1cm}
{\bf {Set V}}
\begin{center}
$ R = \left( \begin{array}{ccc}
   -0.964170&-0.707106 & -0.964171 \\
  1.05385 & 0.0 & 1.05385584\\
 0.161141  & 0.707107  &  0.161141  \end{array} \right)$,\quad~$G^0 = \left( \begin{array}{ccc}
 -0.838569 & 0.0 & 0.0\\
  0.0 & 0.0 & 0.0 \\
  0 & 0  & 0.838569  \end{array} \right)$, ~$ G^1 = \left( \begin{array}{ccc}
 -0.112782 & 0.07359 &  0.143663\\
   0.07359 & -0.104607 & 0.07359 \\
   0.143663 & 0.07359 & 0.143663 \end{array} \right)$.
\end{center}
\vspace{1cm}


\end{document}